\documentclass[aps,twocolumn,superscriptaddress,showpacs]{revtex4-1}
\usepackage{graphicx}
\usepackage{dcolumn}
\usepackage{bm}
\usepackage{amsmath}

\begin{document}

\title{Quantum spin state transitions in spin-1 equilateral triangular lattice antiferromagnet Na$_2$BaNi(PO$_4$)$_2$}

\author{N. Li}
\thanks{These authors contributed equally to this work.}
\affiliation{Department of Physics, Hefei National Laboratory for Physical Sciences at Microscale, and Key Laboratory of Strongly-Coupled Quantum Matter Physics (CAS), University of Science and Technology of China, Hefei, Anhui 230026, People's Republic of China}

\author{Q. Huang}
\thanks{These authors contributed equally to this work.}
\affiliation{Department of Physics and Astronomy, University of Tennessee, Knoxville, Tennessee 37996-1200, USA}

\author{A. Brassington}
\affiliation{Department of Physics and Astronomy, University of Tennessee, Knoxville, Tennessee 37996-1200, USA}

\author{X. Y. Yue}
\affiliation{Institute of Physical Science and Information Technology, Anhui University, Hefei, Anhui 230601, People's Republic of China}

\author{W. J. Chu}
\affiliation{Department of Physics, Hefei National Laboratory for Physical Sciences at Microscale, and Key Laboratory of Strongly-Coupled Quantum Matter Physics (CAS), University of Science and Technology of China, Hefei, Anhui 230026, People's Republic of China}

\author{S. K. Guang}
\affiliation{Department of Physics, Hefei National Laboratory for Physical Sciences at Microscale, and Key Laboratory of Strongly-Coupled Quantum Matter Physics (CAS), University of Science and Technology of China, Hefei, Anhui 230026, People's Republic of China}

\author{X. H. Zhou}
\affiliation{Department of Physics, Hefei National Laboratory for Physical Sciences at Microscale, and Key Laboratory of Strongly-Coupled Quantum Matter Physics (CAS), University of Science and Technology of China, Hefei, Anhui 230026, People's Republic of China}

\author{P. Gao}
\affiliation{Department of Physics, Hefei National Laboratory for Physical Sciences at Microscale, and Key Laboratory of Strongly-Coupled Quantum Matter Physics (CAS), University of Science and Technology of China, Hefei, Anhui 230026, People's Republic of China}

\author{E. X. Feng}
\affiliation{Neutron Scattering Division, Oak Ridge National Laboratory, Oak Ridge, Tennessee 37831, USA}

\author{H. B. Cao}
\affiliation{Neutron Scattering Division, Oak Ridge National Laboratory, Oak Ridge, Tennessee 37831, USA}

\author{E. S. Choi}
\affiliation{National High Magnetic Field Laboratory, Florida State University, Tallahassee, FL 32310-3706, USA}

\author{Y. Sun}
\affiliation{Institute of Physical Science and Information Technology, Anhui University, Hefei, Anhui 230601, People's Republic of China}

\author{Q. J. Li}
\affiliation{School of Physics and Material Sciences, Anhui University, Hefei, Anhui 230601, People's Republic of China}

\author{X. Zhao}
\affiliation{School of Physical Sciences, University of Science and Technology of China, Hefei, Anhui 230026, People's Republic of China}

\author{H. D. Zhou}
\email{hzhou10@utk.edu}
\affiliation{Department of Physics and Astronomy, University of Tennessee, Knoxville, Tennessee 37996-1200, USA}

\author{X. F. Sun}
\email{xfsun@ustc.edu.cn}
\affiliation{Department of Physics, Hefei National Laboratory for Physical Sciences at Microscale, and Key Laboratory of Strongly-Coupled Quantum Matter Physics (CAS), University of Science and Technology of China, Hefei, Anhui 230026, People's Republic of China}
\affiliation{Institute of Physical Science and Information Technology, Anhui University, Hefei, Anhui 230601, People's Republic of China}
\affiliation{Collaborative Innovation Center of Advanced Microstructures, Nanjing University, Nanjing, Jiangsu 210093, People's Republic of China}

\date{\today}

\begin{abstract}

We have grown single crystals of Na$_2$BaNi(PO$_4$)$_2$, a new spin-1 equilateral triangular lattice antiferromagnet (ETLAF), and performed magnetic susceptibility, specific heat and thermal conductivity measurements at ultralow temperatures. The main results are (i) at zero magnetic field, Na$_2$BaNi(PO$_4$)$_2$ exhibits a magnetic ordering at 430 mK with a weak ferromagnetic moment along the $c$ axis. This suggests a canted 120$^\circ$ spin structure, which is in a plane including the crystallographic $c$ axis due to the existence of an easy-axis anisotropy and ferromagnetically stacked along the $c$ axis; (ii) with increasing field along the $c$ axis, a 1/3 magnetization plateau is observed which means the canted 120$^\circ$ spin structure is transformed to a up up down (UUD) spin structure. With even higher fields, the UUD phase further evolves to possible V and V' phases; (iii) with increasing field along the $a$ axis, the canted 120$^\circ$ spin structure is possibly transformed to a umbrella phase and a V phase. Therefore, Na$_2$BaNi(PO$_4$)$_2$ is a rare example of spin-1 ETLAF with single crystalline form to exhibit easy-axis spin anisotropy and series of quantum spin state transitions.

\end{abstract}


\maketitle

\section{Introduction}

The two-dimensional (2D) equilateral triangular lattice antiferromagnet (ETLAF) with spin-1/2 is one of the simplest geometrically frustrated systems with strong quantum spin fluctuations, which recently has caught attention due to its exotic quantum magnetism. One celebrated example is the quantum spin liquid (QSL) \cite{balents2010spin, savary2016quantum, zhou2017quantum, knolle2019field} that can host non-abelian quasiparticle \cite{nayak2008non} and fractional excitations \cite{han2012fractionalized, punk2014topological} known as spinons \cite{kohno2007spinons, nussinov2007high}, which allows quantum mechanical encryption and transportation of information with potential for creating a qubit that is protected against environmental influences \cite{kitaev2006topological}. For example, the recently studied NaYbO$_2$ \cite{bordelon2019field}, NaYbS$_2$ \cite{ma2020spin, PhysRevB.98.220409, PhysRevB.100.241116}, NaYbSe$_2$ \cite{ranjith2019anisotropic, PhysRevB.103.035144, PhysRevX.11.021044}, CsYbSe$_2$ \cite{xing2019field, xie2021}, and YbMgGaO$_4$ \cite{paddison2017continuous, shen2016evidence, li2015gapless, li2017nearest, shen2018fractionalized, li2016muon, li2017spinon, li2017detecting, zhang2018hierarchy} are ETLAFs with effective spin-1/2 for Yb$^{3+}$ ions, and have been proposed as QSL candidates. Due to the Mg/Ga site disorder, the ground state of YbMgGaO$_4$ \cite{xu2016absence, ma2018spin, zhu2017disorder, zhu2018topography, kimchi2018valence, Luo} also has been proposed as spin glass or random spin singlet state.

Another example of exotic magnetism in spin-1/2 ETLAFs is the quantum spin state transition (QSST). The theoretical studies have proposed that in a spin-1/2 ETLAF, the quantum spin fluctuations stabilize a novel up up down (UUD) phase while approaching zero temperature with the applied field parallel to either easy plane or easy axis \cite{chubukov1991quantum, miyashita1986magnetic}. This UUD phase exhibits itself as a magnetization plateau within a certain magnetic field regime and with one-third of the saturation magnetization (1/3 $M_{\text{s}}$). Experimentally, such a UUD phase has been reported for Ba$_3$CoSb$_2$O$_9$ \cite{shirata2012experimental, susuki2013magnetization, zhou2012successive, PhysRevB.103.184425}, Ba$_3$CoNb$_2$O$_9$ \cite{PhysRevB.89.104420, PhysRevB.90.014403}, Ba$_2$La$_2$CoTe$_2$O$_{12}$ \cite{PhysRevB.98.174406}, and Na$_2$BaCo(PO$_4$)$_2$ \cite {li2020possible}, all of which are ETLAFs with effective spin-1/2 for Co$^{2+}$ ions. In some cases the QSL and QSST are correlated. It has been reported that the QSL state can be tuned to the UUD phase by applied magnetic field in these Yb-ETLAFs mentioned above \cite{bordelon2019field, ma2020spin, ranjith2019anisotropic, xing2019field}. Meanwhile, for Na$_2$BaCo(PO$_4$)$_2$, although it orders at $T_{\text{N}}$ = 148 mK, the observed residual term of thermal conductivity ($\kappa_0$/$T$) suggests that it behaves as a QSL with itinerant excitations above its $T_{\text{N}}$ \cite{li2020possible}.

To better understand the quantum magnetism of spin-1/2 ETLAFs, a useful approach is to switch the spin number. For example, one can replace the spin-1/2 by spin-1, and study how the quantum magnetism of the ETLAFs changes accordingly. The reported studies on spin-1 ETLAFs are limited. So far, the UUD phase has been reported for 6HA phase of Ba$_3$NiSb$_2$O$_9$ \cite{shirata2011quantum}, Ba$_3$NiNb$_2$O$_9$ \cite{PhysRevLett.109.257205}, Ba$_2$La$_2$NiTe$_2$O$_{12}$ \cite{PhysRevB.100.064417}, and the QSL has been proposed for 6HB phase of Ba$_3$NiSb$_2$O$_9$ \cite{PhysRevLett.107.197204, PhysRevLett.109.016402, PhysRevB.93.214432, PhysRevB.95.060402}, all of which are spin-1 ETLAFs for Ni$^{2+}$ ions.

In this work, we synthesized a new spin-1 ETLAF Na$_2$BaNi(PO$_4$)$_2$ and grew single-crystal samples to perform DC, AC magnetic susceptibility, specific heat, and thermal conductivity measurements. The results show that it orders at $T_{\text{N}}$ = 430 mK with a weak ferromagnetic moment and exhibits a 1/3 $M_{\text{s}}$ plateau (the UUD phase) with magnetic field applied along the $c$ axis, both of which suggest an easy-axis anisotropy. The obtained magnetic phase diagrams show more field induced spin state transitions besides the UUD phase.

\section{Experiments}

The polycrystalline samples of Na$_2$BaNi(PO$_4$)$_2$ was firstly made by solid state reaction. The stoichiometric mixtures of NaCO$_3$, BaCO$_3$, NiCO$_3$, and (NH$_4$)$_2$HPO$_4$ were ground together, and then annealed in air at 300, 600 and 700 $^\circ$C for 20 hours successively. The single crystal sample was prepared by following the reported procedure used for single crystal growth of Na$_2$BaCo(PO$_4$)$_2$ \cite{Zhong14505}. The polycrystalline sample Na$_2$BaNi(PO$_4$)$_2$ and NaCl were mixed inside a Pt crucible with the mass ratio 1:5. The mixture was heated up to 950 $^\circ$C for 2 hours, and then cooled down to 750 $^\circ$C with a rate of 3 $^\circ$C/hour. The obtained crystals are thin plates with yellow color. The single crystal x-ray diffraction data was collected at 250 K with the Mo K$_{\alpha}$ radiation (0.71073 \AA) using a Rigaku Xtalab Pro diffractormeter at Oak Ridge National Laboratory's neutron scattering user facility. The data reduction was done by the Crysalispro and the refinement was done with the Fullprof \cite{Fullprof}.

DC magnetic susceptibility ($\chi$) was measured with a Quantum Design superconducting quantum interference device (SQUID) magnetometer with a zero field cooling process and a DC magnetic field $B$ = 0.1 T. DC magnetization with magnetic field up to 14 T was measured using a physical properties measurement system (PPMS, Quantum Design). The AC susceptibility ($\chi'$) measurements were conducted with a voltage controlled current source (Stanford Research, CS580) and lock-in amplifier (Stanford Research, SR830) \cite{PhysRevB.89.064401}. The phase of the lock-in amplifier is set to measure the first harmonic signal. The rms amplitude of the ac excitation field is set to be 0.6 Oe with the frequency fixed at 220 Hz. The measurements were performed at SCM1 of the National High Magnetic Field Laboratory, Tallahassee, by using a dilution refrigerator. The data was obtained by the zero field cooling process and we increased the magnetic field during the ramping process.

Specific heat ($C\rm_{p}$) measurement was performed by using the relaxation method on PPMS (Quantum Design) equipped with a dilution refrigerator insert. Thermal conductivity ($\kappa$) was measured using a ``one heater, two thermometers" technique in a $^3$He refrigerator at 300 mK $< T <$ 30 K and a $^3$He/$^4$He dilution refrigerator at 70 mK $< T <$ 1 K, equipped with a 14 T magnet \cite{li2020possible, Rao_YMGO, Song_CFO}. The sample was cut precisely along the crystallographic axes with dimensions of 2.67 $\times$ 0.73 $\times$ 0.14 mm$^3$, where the longest and the shortest dimensions are along the $a$ and $c$ axis, respectively. The heat current was applied along the $a$ axis, while the magnetic field were applied along either the $a$ or $c$ axis.

\section{Results}

\subsection{Crystal structure}

\begin{figure*}[tbp]
\includegraphics[width=18cm]{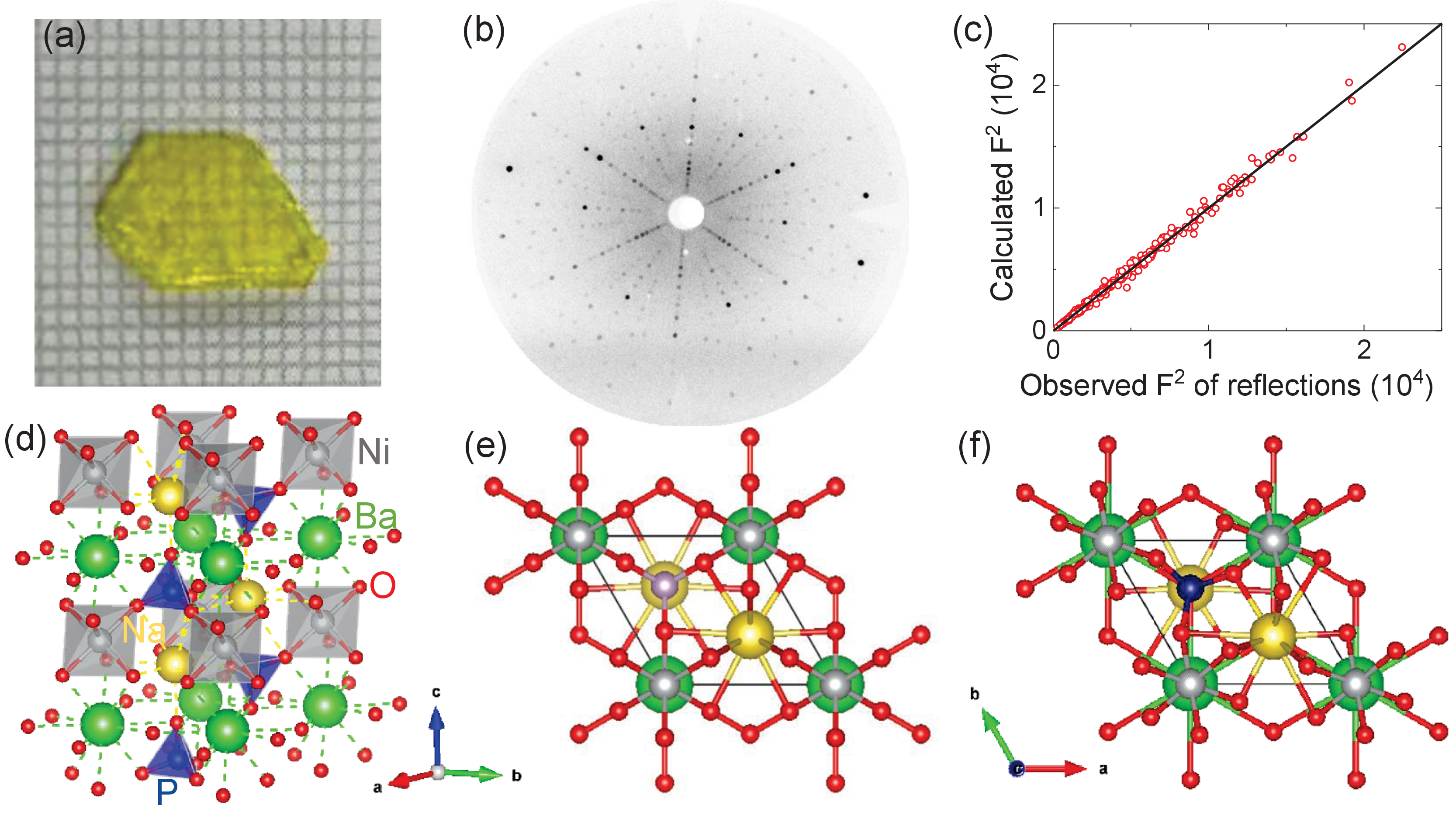}
\caption{(a) As grown single crystal of Na$_2$BaNi(PO$_4$)$_2$. The scale of the grid is 1 mm. (b) The Laue back diffraction pattern along the crystallographic $c$ axis. (c) The comparison between the observed squared structure factors at 250 K by single crystal x-ray diffraction and the calculated ones based on the $P\overline{3}$ structure. (d) The illustration of the crystal structure for Na$_2$BaNi(PO$_4$)$_2$. The projection of the crystal structure along the $c$ axis for Na$_2$BaCo(PO$_4$)$_2$ (e) with $P\overline{3}m1$ and for Na$_2$BaNi(PO$_4$)$_2$ with $P\overline{3}$ (f).}
\end{figure*}

\begin{table}[h]
	\caption{Data of crystallographic refinement for Na$_{2}$BaNi(PO$_4)_2$} 
	\centering 
	\begin{tabular}{c c} 
		\hline\hline 
	    Crystal system & Hexagonal\\ 
		Space group & $P\overline{3}$ \\
		Temperature & 250 K \\
	    Z & 1 \\
		$a$ (\AA) & 5.284(1) \\
		$c$ (\AA) & 6.980(8) \\
		$V$ (\AA$^3$) & 168.801(5) \\
		Extinction Coefficient & 0.2964 \\
		No. reflections; R$_\text{int}$ & 2664; 4.72 \\
		No. independent reflections) & 480 \\
	   R$_\text{F2}$; R$_{\text{F2}\omega}$ & 3.89; 5.10 \\
		R$_\text{F}$; $\chi^2$ & 2.04; 1.12\\

		\hline 
	\end{tabular}
	\label{table:nuclear} 
\end{table}

\begin{table}[h]
	\caption{Atomic positions and parameters of Na$_{2}$BaNi(PO$_4)_2$ at 250K} 
	\centering 
	\begin{tabular}{c c c c c c c} 
		\hline\hline 
		Atom & site & Occ. & x & y & z & U$_\text{eq}$ \\ [0.5ex] 
		\hline 
		Ba & 1a & 1 & 0 & 0 & 0 & 0.0081(2) \\ 
		Ni & 1b & 1 & 0 & 0 & 0.5 & 0.0054(3)\\
		P & 2d & 1 & 1/3 & 2/3 & 0.24466(13) & 0.0054(4) \\
		Na & 2d & 1 & 1/3 & 2/3 & 0.6782(3) & 0.0133(8)\\
		O1 & 2d & 1 & 1/3 & 2/3 & 0.0265(4) & 0.0112(12)\\
		O2 & 6g & 1 & 0.2312(4) & 0.8756(4) & 0.3212(2) & 0.0115(8)\\
		\hline 
	\end{tabular}
	\label{table:nuclear2} 
\end{table}

The as-grown crystals of Na$_2$BaNi(PO$_4$)$_2$ are thin plates with several mm size, as shown in Fig. 1(a). The plate is normal to the crystallographic $c$ axis, as confirmed by the Laue back diffraction pattern (Fig. 1(b)). The best refinement of the single crystal diffraction data (Fig. 1(c)) leads to a $P\overline{3}$ space group. The detailed crystallographic parameters and atomic positions are listed in Table I and II, respectively. The crystal structure is shown in Fig. 1(d), in which the triangular layers of magnetic NiO$_6$ octahedra are separated by a single layer of nonmagnetic BaO$_{12}$ polyhedra, with [PO$_4$]$^{3-}$ units and Na$^+$ filling the gaps in the nickel oxide layers. This structure is very similar to that of Na$_2$BaCo(PO$_4$)$_2$ with $P\overline{3}m1$ space group but with some difference. Compared to the $P\overline{3}m1$ space group (shown in Fig. 1(e)), in $P\overline{3}$ space group (Fig. 1(f)), the nonmagnetic BaO$_{12}$ polyhedra are fixed but the NiO$_6$ octahedra and PO$_4$ tetrahedra rotate in the opposite direction along the $c$ axis (Fig. 1(f)). Therefore, the mirror plane symmetry in Na$_2$BaCo(PO$_4$)$_2$ is eliminated in Na$_2$BaNi(PO$_4$)$_2$.

\subsection{DC and AC susceptibility}

\begin{figure}
\includegraphics[clip,width=8.5cm]{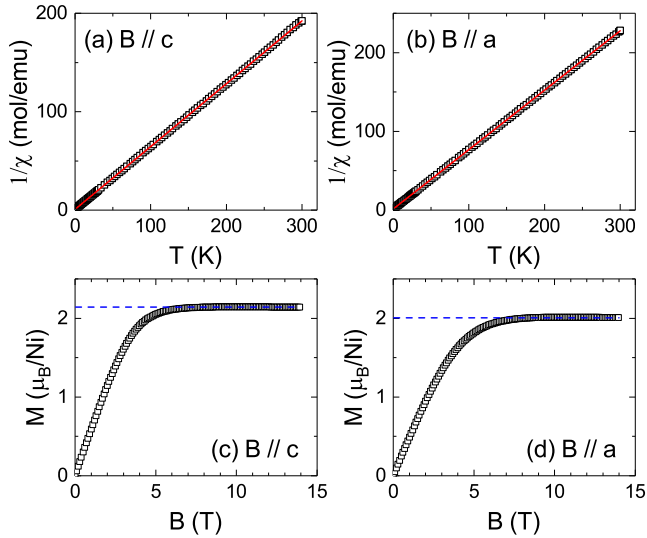}
\caption{Temperature dependency of the inverse DC magnetic susceptibility with $B \parallel c$ (a) or $B \parallel a$ (b). The DC magnetization measured at 2 K with $B \parallel c$ (c) or $B \parallel a$ (d). The solid lines in (a,b) are linear Curie-Weiss fittings. The dashed lines in (c,d) indicate the spin saturation.}
\label{ACT}
\end{figure}

Figures 2(a) and 2(b) show the temperature dependence of the inverse DC susceptibility, 1/$\chi$, for $B \parallel c$ and $B \parallel a$, respectively. The data shows no sign of magnetic ordering down to 2 K. and can be well fitted by the Curie-Weiss law with $\mu_{\text{eff}}$ = 3.56(1) $\mu_{\text{B}}$, $\theta_{\text{CW}} = -1.2(1)$ K for $B \parallel c$ and $\mu_{\text{eff}}$ = 3.26(1) $\mu_{\text{B}}$, $\theta_{\text{CW}} = -1.9(1)$ K for $B \parallel a$. Figures 2(c) and 2(d) show the DC magnetization measured at 2 K for $B \parallel c$ and $B \parallel a$, respectively. The saturation moments are 2.14(5) $\mu_{\text{B}}$ for $B \parallel c$ and 2.00(5) $\mu_{\text{B}}$ for $B \parallel a$. The $g$ factors parallel and perpendicular to the $c$ axis were evaluated to be $g_{\parallel}$ = 2.14(5) and $g_{\perp}$ = 2.00(5).

\begin{figure}
\includegraphics[clip,width=8.5cm]{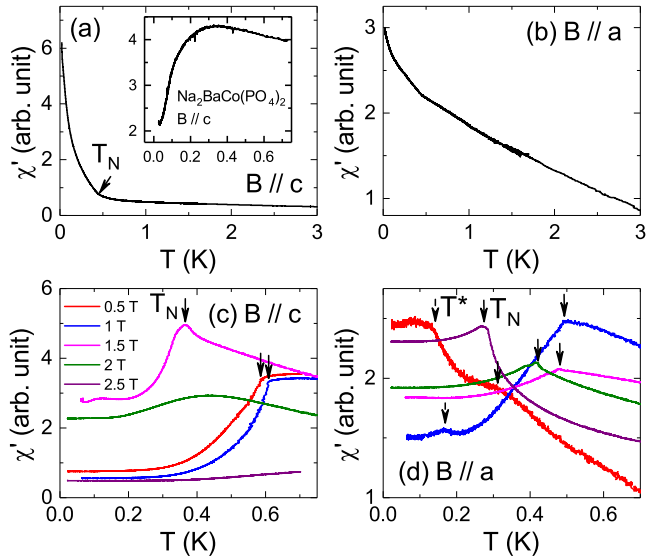}
\caption{Temperature dependence of the AC magnetic susceptibility measured at zero DC field but with the AC excitation field along the $c$ axis (a) or $a$ axis (b), or at different DC fields along the $c$ axis (c) or $a$ axis (d). The solid arrows represent $T_{\text{N}}$ and the dashed arrows represent $T$*. For comparison, the inset shows the AC susceptibility data of Na$_2$BaCo(PO$_4$)$_2$ single crystal with the AC excitation field along the $c$ axis.}
\label{ACT2}
\end{figure}

Figure 3(a) shows the temperature dependence of the AC susceptibility, $\chi'(T)$, measured at zero DC magnetic field ($B$ = 0 T) but with the AC excitation field along the $c$ axis. The curve shows a rapid increase below $\sim$400 mK (defined as N\'eel temperature $T_{\text{N}}$), indicating a magnetic ordering, similar to that observed in some other ETLAFs
like Ba$_2$La$_2$NiTe$_2$O$_{12}$ \cite{PhysRevB.100.064417}. For the data measured with the AC excitation field along the $a$ axis (Fig. 3(b)), this feature is not so clear. With applied magnetic field along both the $c$ and $a$ axes, $\chi'(T)$ shows a peak at $T_{\text{N}}$ (Figs. 3(c) and 3(d)). With increasing field, this peak first shifts to higher temperatures while $B \leq$ 1 T, and then comes back to lower temperatures while 1 $< B \leq$ 2.5 T. For the data measured with $B \parallel a$ and $B$ = 0.5 and 1.0 T, another additional peak is observed below $T_{\text{N}}$, which is labeled as $T$* in Fig. 3(d).

\begin{figure}
\includegraphics[clip,width=8.5cm]{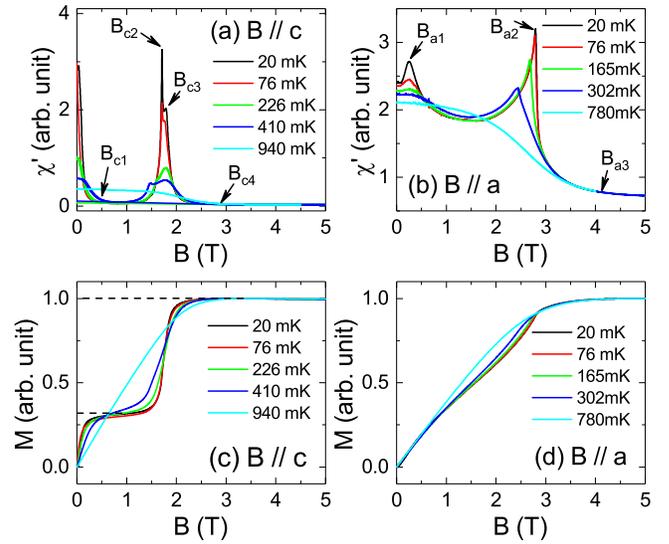}
\caption{DC field dependence of the AC magnetic susceptibility at different temperatures with $B \parallel c$ (a) and $B \parallel a$ (b). Magnetization obtained by integrating the AC susceptibility data for $B \parallel c$ (c) and $B \parallel a$ (d).}
\label{ACH}
\end{figure}

Figure 4(a) shows the field dependence of $\chi'(B)$ measured at different temperatures with $B \parallel c$. At 20 mK, with increasing field, the $\chi'(B)$ first quickly drops to a flat regime at $B_{\text{c1}}$ = 0.32(2) T, and then exhibits two peaks at $B_{\text{c2}}$ = 1.72(1) T and $B_{\text{c3}}$ = 1.80(1) T, and finally drops to be flat again at $B_{\text{c4}}$ = 2.84(2) T. With increasing temperature, $B_{\text{c1}}$ and $B_{\text{c2}}$ shift to higher and lower fields respectively while $B_{\text{c3}}$ and $B_{\text{c4}}$ show no obvious change. Meanwhile, the peaks around $B_{\text{c2}}$ and $B_{\text{c3}}$ become broader with increasing temperature and disappear with temperature above 410 mK. Since the measured $\chi'$($B$) shows no frequency dependence (not shown here), it could be approximately treated as the derivative of the DC magnetization $M(B)$. Thus, we calculated $M(B)$ by integrating $\chi'(B)$. The obtained $M(B)$ at 20 mK (Fig. 4(c)) clearly shows a plateau regime between $B_{\text{c1}}$ and $B_{\text{c2}}$ and a slope change at $B_{\text{c3}}$ followed by the saturation around $B_{\text{c4}}$. Although we cannot infer the absolute value of $M(B)$ here, it is obvious that the magnetization of the plateau (around 0.32, here we scaled the $M$ value to the 5.0 T value) is around 1/3 of the saturation value (1.0).

Figure 4(b) shows the $\chi'(B)$ with $B \parallel a$. At 20 mK, the curve shows two peaks at $B_{\text{a1}}$ = 0.230(2) T and $B_{\text{a2}}$ = 2.80(2) T, respectively, and then drops to be flat at $B_{\text{a3}}$ = 4.00(5) T. With increasing temperature, both peaks become broader and the $B_{\text{a1}}$ peak shifts to higher fields and the $B_{\text{a2}}$ peak shifts to lower fields. Above 300 mK, the peaks disappear. The calculated $M(B)$ at 20 mK with $B \parallel a$ (Fig. 4(d)) shows no obvious sign for the magnetization plateau, which is clearly different from that for $B \parallel c$.

\subsection{Specific heat}

\begin{figure}
\includegraphics[clip,width=6cm]{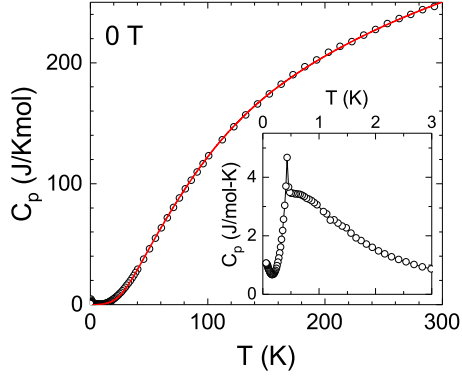}
\caption{Temperature dependence of the specific heat measured at zero field. The red line is the fitting of phonon specific heat as described in the main text. The inset highlights the low temperature specific heat data.}
\label{Cp}
\end{figure}

Figure 5 shows the specific heat, $C\rm_{p}$, measured at zero field. As shown in the inset, the low-temperature data exhibits a broad peak around 600 mK followed by a small but sharp peak at $T_{\text{N}}$ = 430(3) mK. This sharp peak indicates a long-range magnetic ordering, which is consistent with the rapid increase at $\sim$400 mK in the zero-field $\chi'(T)$ curve. Apparently, the phonon specific heat is dominant at high-temperature regime from several tens to 300 K. However, it was found that the standard Debye formula cannot fit the high-temperature at all. One known reason for the deviations of high-$T$ specific heat from the Debye model is the contribution of optical phonons at high temperatures, which can be described by the Einstein model \cite{Song_CFO, Svoboda, Hemberger, Janiceka}. The phonon spectrum of Na$_2$BaNi(PO$_4$)$_2$ should consist of 3 acoustic branches and 39 optical branches. We found that the high-temperature data can be fitted by the formula
\begin{equation}\label{eq:eps}
\begin{split}
C_{\text{ph}} = & \ 3N_D R \left(\frac{T}{\Theta_D}\right)^3 \int_0^{\Theta_D /T} \frac{x^4e^x}{(e^x-1)^2}\mathrm{d}x \\
& + N_{E1} R \left(\Theta_{E1}/T\right)^2 \frac{\mathrm{exp}(\Theta_{E1}/T)}{[\mathrm{exp}(\Theta_{E1}/T)-1]^2} \\
& + N_{E2} R \left(\Theta_{E2}/T\right)^2 \frac{\mathrm{exp}(\Theta_{E2}/T)}{[\mathrm{exp}(\Theta_{E2}/T)-1]^2} \\
& + N_{E3} R \left(\Theta_{E3}/T\right)^2 \frac{\mathrm{exp}(\Theta_{E3}/T)}{[\mathrm{exp}(\Theta_{E2}/T)-1]^2},
\end{split}
\end{equation}
with $x = \hbar\omega/k_BT$, $\omega$ the phonon frequency, $k_B$ the Boltzmann constant, and $R$ the universal gas constant. Here, the first term is the contribution of 3 acoustic phonon branches using the Debye model ($N_D =$ 3), while the other terms are the contributions from the optical branches using the Einstein model ($N_{E1} =$ 12, $N_{E2} =$ 15 and $N_{E3} =$ 12). Other parameters are the Debye temperature, $\Theta_D =$ 165 K, and three Einstein temperatures, $\Theta_{E1} =$ 202 K, $\Theta_{E2} =$ 451 K and $\Theta_{E3} =$ 1285 K.

\begin{figure}
\includegraphics[clip,width=8.5cm]{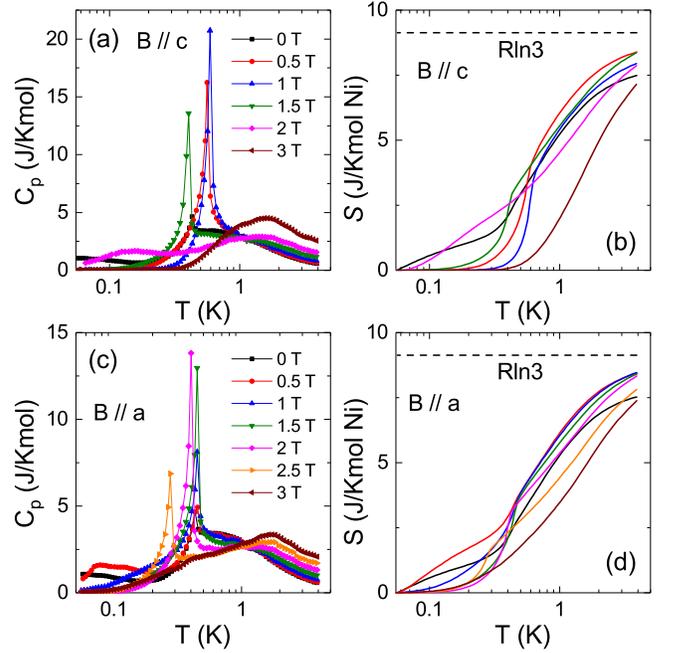}
\caption{(a,c) Specific heat data with $T <$ 4 K and in different magnetic fields along the $c$ axis or the $a$ axis. (b,d) The magnetic entropy for $B \parallel c$ and $B \parallel a$, obtained by integrating the magnetic specific heat. The dashed lines show the entropy for $R$ln3.}
\label{Cp2}
\end{figure}

Figures 6(a) and 6(c) show the $C\rm_{p}$ data measured at $T <$ 4 K and in different fields. With increasing field along the $c$ axis, the peak at $T_{\text{N}}$ first shifts to higher temperatures with abruptly increased intensity while $B < $ 1 T, then comes back to lower temperatures with decreased intensity while 1 $<$ $B$ $<$ 2 T, and disappears while $B \geq$ 2 T. Meanwhile, the broad peak above $T_{\text{N}}$ shifts to higher temperatures around 1 $\sim$ 2 K and gains intensity with increasing field. The change of the magnetic entropy below 4 K, $\Delta S_{\text{m}}$, was calculated by integrating ($C_{\text{p}} - C_{\text{ph}})/T$ (Fig. 6(b)). At zero field, the obtained value is 7.46 J/Kmol Ni, which is approaching the expected magnetic entropy value of a spin-1 system, $\Delta S_{\text{m}}$ = $R$ln(2$S$+1) = $R$ln3 with $S$ = 1. The recovered magnetic entropy below 480 mK (where the sharp peak starts) is 2.83 J/Kmol Ni, which is only 31\% of $R$ln3. This indicates a large portion of the magnetic entropy has already been recovered or the existence of strong spin fluctuations above the long range magnetic ordering temperature, $T_{\text{N}}$. The broad peak around 600 mK  at zero field then could be due to the development of short-range correlations. Accordingly, the shift to higher temperatures with increased intensity for this broad peak under fields suggests that applied field enhances the short-range correlations. Meanwhile, the applied field tunes the long-range ordering in a non-monotonic way. Similar behaviors were observed for the $C\rm_{p}$ measured with $B \parallel a$, as shown in Figs. 6(c) and 6(d).

\subsection{Thermal conductivity}

\begin{figure}
\includegraphics[clip,width=6.5cm]{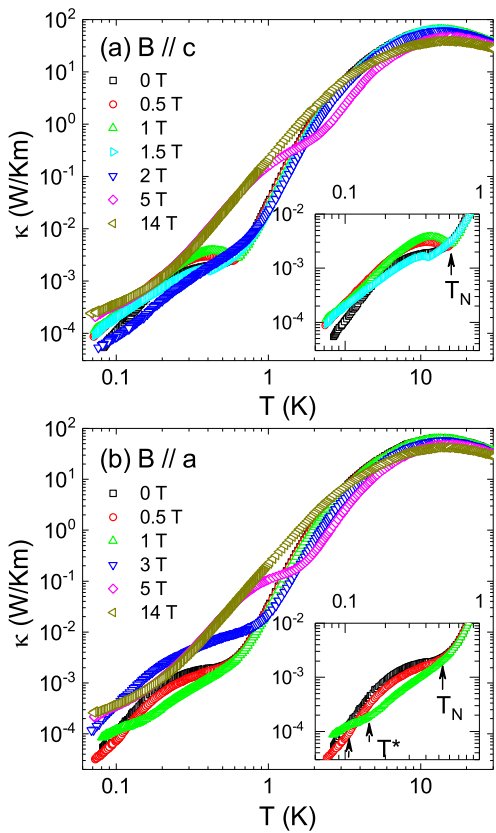}
\caption{Temperature dependence of the thermal conductivity measured at different magnetic fields along the $c$ axis (a) or the $a$ axis (b). The insets of (a) and (b) highlight the low temperature data. The solid arrows represent $T_{\text{N}}$ and the dashed arrows represent $T$*.}
\label{kappaT}
\end{figure}

Figure 7 shows the temperature dependence of $\kappa$ measured at different magnetic fields. The zero-field thermal conductivity exhibits a phonon peak at about 14 K with a rather large value of 65 W/Km, which indicates the high crystalline quality of the Na$_2$BaNi(PO$_4$)$_2$ sample. In addition, there is a kink-like feature around 420 mK, which again should be related to the magnetic ordering revealed by both the $\chi'(T)$ and $C\rm_{p}$. With increasing field along the $c$ axis, as shown in the inset of Fig. 7(a), the kink changes to a dip at 0.5 and 1 T and the dip position shifts to higher temperatures at 582(5) mK and 612(5) mK, respectively. With $B =$ 1.5 T the dip becomes weaker and shifts to lower temperature of 429(5) mK and then disappears with $B \geq$ 2 T. This evolution of $T_{\text{N}}$ under fields is similar to that observed from the $\chi'(T)$ and $C\rm_{p}$ data with $B \parallel c$. With $B =$ 2 T and 5 T, a new kink appears around 640 mK and 1.9 K, respectively. This feature is likely associated with the phonon scattering by spin fluctuations since the $C\rm_{p}$ data measured at 2 T and 3 T indicate that the higher fields enhance the short-range correlations around 1 $\sim$ 2 K. With increasing field along the $a$ axis, as shown in Fig. 7(b), the zero-field kink gradually moves to higher temperatures and disappears with $B\geq$ 3 T, while the 5 T and 14 T data are almost the same as those for $B \parallel c$. It is noticed that the 0.5 T and 1 T data with $B \parallel a$ shows an additional kink around $T$* = 110(2) and 160(3) mK, respectively (inset of Fig. 7(b)), which are consistent with the $T$* observed from the $\chi'(T)$ data.

\begin{figure}
\includegraphics[clip,width=8.5cm]{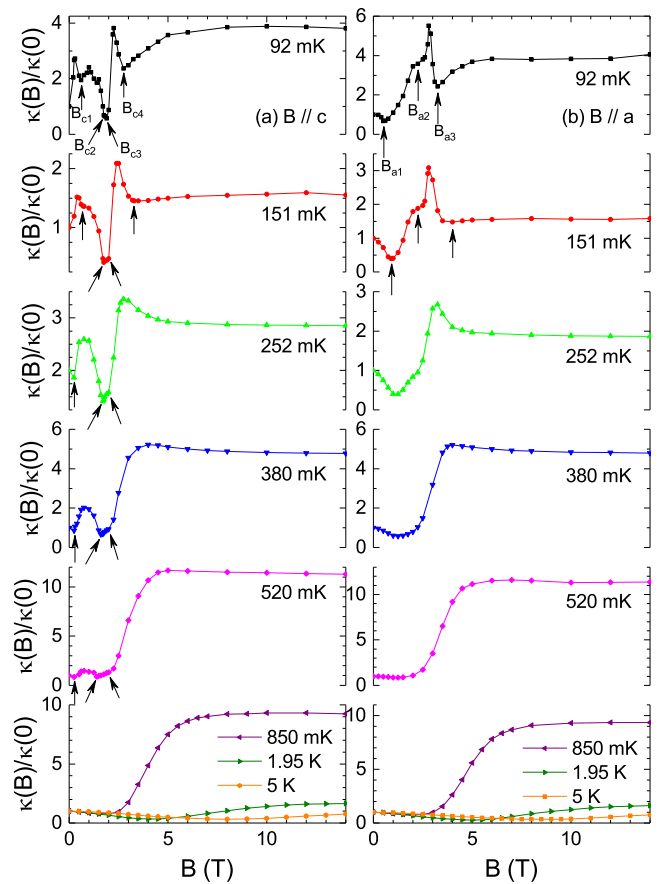}
\caption{Thermal conductivity as a function of the magnetic field along the $c$ axis (a) and the $a$ axis (b).}
\label{kappaH}
\end{figure}

It is well known from many previous studies on the low-dimensional or frustrated spin systems that the anomalies, either the dip- or kink-like features, on the $\kappa(T)$ curves are usually related to the magnetic transitions. In addition, the field dependence of $\kappa$ can also display similar anomalies at the field-induced magnetic transitions \cite{PhysRevB.103.184425, li2020possible, Rao_YMGO, Song_CFO, Leahy, Hentrich, Buys}. Figure 8 shows the $\kappa(B)$ curves at low temperatures for Na$_2$BaNi(PO$_4$)$_2$ single crystal. At 92 mK, the curve with $B \parallel c$ exhibits four minimums at $B_{\text{c1}}$, $B_{\text{c2}}$, $B_{\text{c3}}$, and $B_{\text{c4}}$ while there are three minimums at $B_{\text{a1}}$, $B_{\text{a2}}$, and $B_{\text{a3}}$ for the $B \parallel a$ curve. The values of these critical fields are consistent with those observed from the $\chi'(B)$ data. It is noticed that with increasing temperature, the separation between $B_{\text{c2}}$ and $B_{\text{c3}}$ increases, which is also observed from the $\chi'(B)$ data. Above certain temperatures, such as 850 mK for $B \parallel c$ and 380 mK for $B \parallel a$, these critical fields are not recognizable.

Another feature of the low-temperature $\kappa(B)$ curves is that the $\kappa$ at high field is always larger than the zero-field $\kappa$ and shows saturation behavior. This means that in zero field there is significant magnetic scattering of phonons in a broad temperature region, including both $T < T_{\text{N}}$ and $T > T_{\text{N}}$. In particular, at $T > T_{\text{N}}$ like 520 and 850 mK, the strong recovery of $\kappa$ at high field indicates the existence of strong spin fluctuations in zero field that can scatter phonons.

\subsection{Phase diagram}

\begin{figure}
\includegraphics[clip,width=6cm]{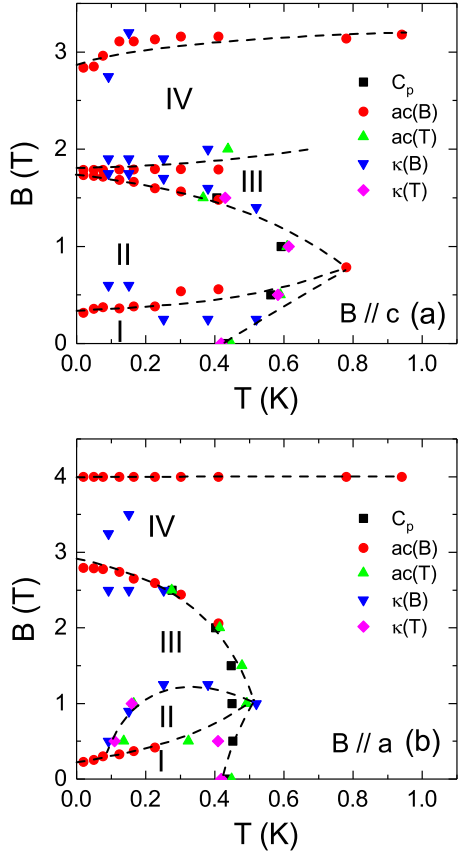}
\caption{Magnetic phase diagrams of Na$_2$BaNi(PO$_4$)$_2$ for $B \parallel c$ (a) and $B \parallel a$ (b).}
\label{phase diagram}
\end{figure}

The consistent behaviors for the field dependence of critical temperatures ($T_{\text{N}}$, $T$*) and temperature dependence of all critical fields among the $\chi'$, $C\rm_{p}$, and $\kappa$ data strongly suggest the occurrence of field-induced spin state transitions. Accordingly, the magnetic phase diagrams for $B \parallel c$ and $B \parallel a$ were constructed, as shown in Fig. 9. Besides the paramagnetic phase at high temperatures and fully polarized phase at high fields, there are four phases for both cases.

\section{Discussions}

Since at low frequency the temperature dependence of the AC $\chi'(T)$ should behave similarly to that of its DC $\chi(T)$, the rapid increase of $\chi'(T)$ below $T_{\text{N}}$ with $B \parallel c$ for Na$_2$BaNi(PO$_4$)$_2$ indicates a magnetic ground state with a weak ferromagnetic moment along the $c$ axis. Moreover, the rapid increase of $M(B)$  at low fields (the rapid decrease or equally a large peak near 0 T for $\chi'(B)$) also could be seen as a supportive feature for the weak ferromagnetism. This behavior is different from that of several studied ETLAFs. As shown in the inset of Fig. 3(a), the $\chi'(T)$ with $B \parallel c$ for Na$_2$BaCo(PO$_4$)$_2$ shows a broad peak at 300 mK followed by a kink at $T_{\text{N}}$ = 148 mK \cite{li2020possible}. The $\chi(T)$ for the well studied Ba$_3$CoSb$_2$O$_9$ also exhibits a broad peak at 7.0 K followed by an anomaly at $T_{\text{N}}$ = 3.6 K \cite{zhou2012successive}. The neutron diffraction data of Ba$_3$CoSb$_2$O$_9$ confirmed a 120$^\circ$ spin structure \cite{PhysRevLett.116.087201}. While such kind of 120$^\circ$ spin structure is common for ETLAFs, it does not lead to a net ferromagnetic moment and therefore should not be the ground state for Na$_2$BaNi(PO$_4$)$_2$.

Meanwhile, the reported Ba$_2$La$_2$NiTe$_2$O$_{12}$ exhibits a similar rapid increase of $\chi(T)$ below $T_{\text{N}}$ \cite{PhysRevB.100.064417}. Its neutron powder diffraction data confirmed a Y-like triangular spin structure, or a canted 120$^\circ$ spin structure, which is in a plane including the crystallographic $c$ axis due to the existence of an easy-axis  anisotropy and ferromagnetically stacked along the $c$ axis. In this canted 120$^\circ$ spin structure, the angle $\theta$ between the canted sublattice spins and the $c$ axis is smaller than 60$^\circ$. Accordingly, the sum of the magnetic moments of the three sublattice spins is nonzero but lead to a resultant magnetic moment along the $c$ axis in a triangular layer. Now if the triangular layers are ferromagnetically stacked, all the resultant magnetic moments appearing in the triangular layers will align in the same direction and result in a net magnetic moment along the $c$ axis. Here, we propose that Na$_2$BaNi(PO$_4$)$_2$ also has such a magnetic ground state, a canted 120$^\circ$ spin structure ferromagentically stacked along the $c$ axis with easy-axis anisotropy. Accordingly, we tend to assign the Phase I for both $B \parallel c$ and $B \parallel a$ to be the canted 120$^\circ$ spin state. While the exact reason for the different magnetic ground states between Na$_2$BaNi(PO$_4$)$_2$ and Na$_2$BaCo(PO$_4$)$_2$ is not clear, it could be due to the lower symmetry in lattice structure for Na$_2$BaNi(PO$_4$)$_2$.

\begin{figure}
\includegraphics[clip,width=8.5cm]{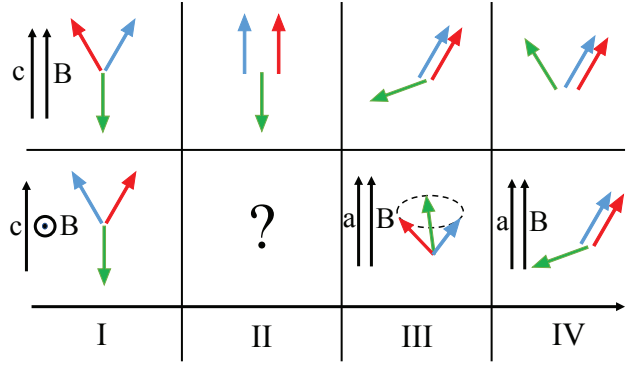}
\caption{The proposed spin structure in each phase for $B \parallel c$ (upper row) and $B \parallel a$ (bottom row).}
\label{kappaH}
\end{figure}

It is obvious that the Phase II for $B \parallel c$ is the UUD phase because for this phase its magnetization is a 1/3 $M_{\text{s}}$ plateau. Since the UUD phase only survives for $B \parallel c$ in Na$_2$BaNi(PO$_4$)$_2$, it again suggests its easy-axis anisotropy as the theory predicted \cite{miyashita1986magnetic}. We can further compare the magnetic phase diagrams of Na$_2$BaNi(PO$_4$)$_2$ to those of Ba$_3$CoSb$_2$O$_9$. For Ba$_3$CoSb$_2$O$_9$ with easy-plane anisotropy, with increasing field along the $ab$ plane, its 120$^\circ$ spin structure at zero field is followed by a canted 120$^\circ$ spin structure, the UUD phase, a coplanar phase (the V phase), and another coplanar phase (the V') phase before entering the fully polarized state  \cite{PhysRevB.103.184425, PhysRevLett.112.127203, PhysRevB.93.224402, PhysRevB.91.081104, PhysRevB.91.024410, PhysRevB.94.214408, PhysRevB.100.094436}. While for $B \parallel c$, the 120$^\circ$ spin structure will be followed by an umbrella spin structure, and the V phase. Since Na$_2$BaNi(PO$_4$)$_2$ has easy-axis anisotropy, we made an analogy between its $B \parallel c$ phase diagram and the $B \parallel a$ phase diagram of Ba$_3$CoSb$_2$O$_9$ or its $B \parallel a$ phase diagram and the $B \parallel c$ phase diagram of Ba$_3$CoSb$_2$O$_9$. By doing that, we tend to assign the phase III and IV with $B \parallel c$ for Na$_2$BaNi(PO$_4$)$_2$ to be the V and V' phases, respectively, and the phase III and IV with $B \parallel a$ to be the umbrella and V phases, respectively. Certainly, further studies such as neutron diffraction measurements are needed to verify the spin structures of these phases. Moreover, it is surprising to see a phase II with $B \parallel a$ that is surrounded by phase III. This phase could be a metastable spin state between the canted 120$^\circ$ and the umbrella phase, whose nature again needs future studies to be clarified. The proposed spin structure in each phase is summarized in Fig. 10.

Finally, we want to point out that, to our knowledge, Na$_2$BaNi(PO$_4$)$_2$ is a rare example of spin-1 ETLAF with single crystalline form to exhibit series of QSSTs. The studied other spin-1 ETLAFs with Ni$^{2+}$ ions, including Ba$_3$NiSb$_2$O$_9$, Ba$_3$NiNb$_2$O$_9$, and Ba$_2$La$_2$NiTe$_2$O$_{12}$ are all with polycrystalline form. The availability of single crystalline Na$_2$BaNi(PO$_4$)$_2$ makes it an excellent candidate for future spin-1 ETLAF studies, such as the inelastic neutron scattering measurements. For example, for Ba$_3$CoSb$_2$O$_9$, unusual spin dynamics have been confirmed by the inelastic neutron scattering experiments \cite{zhou2012successive, PhysRevLett.116.087201, ISI:000407401000006, ISI:000438031000004, PhysRevB.102.064421, ISI:000478082100017}, including the strong downward re-normalization of the magnon dispersion, roton-like minima, the line broadening throughout the entire Brillouin zone, and the intense dispersive excitation continua extending to a high energy six times of the exchange constant. Therefore, it will be interesting to see how these features can be affected in a spin-1 ETLAF system, which will provide useful information to better understand the role of quantum spin fluctuations in ETLAFs.

\section{SUMMARY}

We have grown single crystals of a new spin-1 ETALF, Na$_2$BaNi(PO$_4$)$_2$, and studied its magnetic susceptibility, specific heat and thermal conductivity at ultralow temperatures. The main experimental results include: (i) this material shows a magnetic ordering at 430 mK with a weak ferromagnetic moment along the $c$ axis; (ii) it exhibits a 1/3 magnetization plateau with a magnetic field applied along the $c$ axis; (iii) there are successive QSSTs for either $B \parallel c$ or $B \parallel a$. It is a rare example of spin-1 ETLAF with single crystalline form to exhibit easy-axis spin anisotropy and series of QSSTs.

\begin{acknowledgements}

This work was supported by the National Natural Science Foundation of China (Grant Nos. U1832209, 11874336, and 11904003), the Nature Science Foundation of Anhui Province (Grant No. 1908085MA09), and the Innovative Program of Hefei Science Center CAS (Grant No. 2019HSC-CIP001).  The work at the University of Tennessee (Q.H., A. B., and H. D.Z.) was supported by NSF with  Grant No. NSF-DMR-2003117. The work performed in NHMFL was supported by NSF-DMR-1157490 and the State of Florida. E.F. and H.B.C. acknowledge the support of U.S. DOE BES Early Career Award No. KC0402020 under Contract No. DE-AC05-00OR22725.

\end{acknowledgements}

\end{document}